\documentclass[aps,prd,nofootinbib,superscriptaddress,prd,aps,floatfix]{revtex4}
\usepackage[latin9]{inputenc}
\usepackage{color,textcomp,amsmath,amssymb,graphicx,amsfonts,graphics,epstopdf,slashed,multirow,adjustbox,lipsum,rotating,xcolor,wrapfig,epsfig}
\usepackage[unicode=true, bookmarks=false, breaklinks=false,pdfborder={0 0 1},backref=section,colorlinks=false]{hyperref}
\hypersetup{colorlinks,bookmarksopen,bookmarksnumbered,linkcolor=blu,pdfstartview=FitH,urlcolor=rossos,citecolor=rossos}
\usepackage[justification=raggedright, margin=5mm,font={sf,small},labelfont=bf, format=plain]{caption}

\def\simlt{\stackrel{<}{{}_\sim}}
\def\simgt{\stackrel{>}{{}_\sim}}

\definecolor{rosso}{cmyk}{0,1,1,0.4}
\definecolor{rossos}{cmyk}{0,1,1,0.55}
\definecolor{rossoc}{cmyk}{0,1,1,0.2}
\definecolor{blu}{cmyk}{1,1,0,0.3}
\definecolor{blus}{cmyk}{1,1,0,0.6}
\definecolor{bluc}{cmyk}{1,1,0,0.1}
\definecolor{verde}{cmyk}{0.92,0,0.59,0.25}
\definecolor{verdec}{cmyk}{0.92,0,0.59,0.15}
\definecolor{verdes}{cmyk}{0.92,0,0.59,0.4}

\begin{document}

\title{Gamma rays from Dark Matter Annihilation in Three-loop Radiative Neutrino Mass Generation Models}

\author{Talal Ahmed Chowdhury}
\email{talal@du.ac.bd}
\affiliation{Department of Physics, University of Dhaka, P.O. Box 1000,
Dhaka, Bangladesh.}
\affiliation{The Abdus Salam International Centre for Theoretical Physics, Strada
Costiera 11, I-34014, Trieste, Italy.}

\author{Salah Nasri}
\email{snasri@uaeu.ac.ae}
\affiliation{Department of Physics, UAE University, P.O. Box
17551, Al-Ain, United Arab Emirates}
\affiliation{The Abdus Salam International Centre for Theoretical Physics, Strada
Costiera 11, I-34014, Trieste, Italy.}

\begin{abstract}
We present the Sommerfeld enhanced Dark Matter (DM) annihilation into gamma ray  for a class of three-loop radiative neutrino mass models with large electroweak multiplets where the DM mass is in O(TeV) range. We show that in this model, the DM annihilation rate becomes more prominent for larger multiplets and it is already within the reach of currently operating Imaging Atmospheric Cherenkov telescopes (IACTs), High Energy Stereoscopic System (H.E.S.S.). Furthermore, Cherenkov Telescope Array (CTA), which will begin operating in 2030, will improve this sensitivity by a factor of $\mathcal{O}{(10)}$ and may exclude a large portion of parameter space of this radiative neutrino mass model with larger electroweak multiplet. This implies that the only viable option is the model with lowest electroweak multiplets i.e. singlets of $SU(2)_{L}$ where the DM annihilation rate is not Sommerfeld enhanced and hence it is not yet constrained by the indirect detection limits from H.E.S.S. or future CTA.
\end{abstract}

\maketitle

\section{Introduction}\label{intro}

We are yet to identify the mass and particle nature of the Dark Matter (DM) of the universe despite having extensive astrophysical and cosmological observations supporting its existence.
Recently the DM, having quantum numbers under the Standard Model (SM) gauge group $SU(2)_{L}\times U(1)_{Y}$ and with mass in the TeV range, has come under the focus of Imaging Atmospheric Cherenkov Telescopes (IACT) as gamma ray produced in the DM annihilation at the central region of the Milky Way galaxy is within the detection reach of IACTs
\cite{Bringmann:2012ez, Wood:2013taa, Buckley:2013bha, Cirelli:2015gux}. 

The flux of the gamma ray photons from cosmic sources rapidly falls when their energies reach  $E\simgt \text{O(TeV)}$. For this reason, the satellite which has the detection area of sub-$\text{m}^{2}$, is not very sensitive for the very high energetic (VHE) gamma ray of the energy range, 100 GeV - 100 TeV. On the other hand, such high energy gamma ray interacts with the upper region of the atmosphere and creates a shower of very energetic secondary charged particles which reaches at about 10 km height. These particles move faster than the speed of light in the air and therefore emit the faint blue Cherenkov light. This Cherenkov light is beamed around the direction of the incident primary photon and it illuminates the ground of about $50000\,\text{m}^{2}$, which is often referred as Cherenkov light pool. Only 100 photons per $\text{m}^{2}$ on the ground can be seen for a primary photon of TeV energy. So if a telescope is somewhere within the light pool and has large mirror area to collect enough photons, it will observe the air shower. Therefore, the effective detection area of a Cherenkov telescope is approximately given by the area of the light pool which is much larger than that of a satellite. The prominent IACTs like MAGIC \cite{magic}, VERITAS \cite{veritas}, CANGAROO \cite{cangaroo} and H.E.S.S. \cite{hess} have revealed intriguing astrophysical VHE gamma ray sources of our universe. In \cite{::2016jja, Abdalla:2016olq} H.E.S.S. collaboration presented search result for gamma signal coming from DM annihilation in the inner region of Milky Way and put upper limits on the annihilation cross section $\langle \sigma v \rangle$ for the DM in the TeV mass range that is not within the reach of collider searches or direct detection experiments. 

In addition, the non-relativistic (NR) DM which has O(TeV) mass and electroweak charge, receives the non-perturbative Sommerfeld enhancement \cite{sommerfeldref, Hisano:2002fk, Hisano:2003ec, Hisano:2004ds, Hisano:2005ec, Hisano:2006nn, Cirelli:2007xd, ArkaniHamed:2008qn, Lattanzi:2008qa, Pieri:2009zi, Iengo:2009ni, Cassel:2009wt, Slatyer:2009vg, Feng:2009hw, Buckley:2009in, Feng:2010zp, Hryczuk:2010zi, Hryczuk:2011vi, McDonald:2012nc, Tulin:2013teo, Cohen:2013ama, Ovanesyan:2014fwa, Baumgart:2014vma, Beneke:2014gja} which increases the DM annihilation rate into gauge bosons, i.e $\text{DM}\,\,\text{DM}\rightarrow WW,ZZ,\gamma\gamma,\gamma Z$ ,significantly. As a consequence, this class of NR DM with TeV mass range, has better prospects of being detected in currently operating H.E.S.S or future CTA \cite{cta, Acharya:2017ttl} which will begin operation in about 2030.

Apart from the DM nature of the universe, the origin and smallness of the neutrino mass is yet to be concluded. The Krauss-Nasri-Trodden (KNT) model \cite{Krauss:2002px} ties these two issues together by radiatively generating neutrino mass\footnote{For a review of radiative neutrino mass generation models, please see \cite{Cai:2017jrq}.} at three-loop with DM particle running in the loop. The Beyond Standard Model (BSM) content of the model consists of two single charged singlet scalars, $S^{+}_{1},\,S^{+}_{2}$ and three singlet RH neutrinos, $N_{R_{i}},\,i=1,2,3$ under SM gauge group with masses lie in the GeV-TeV range. Here, the lightest singlet RH neutrino $N_{R_{1}}$ plays the role of DM. Subsequently, KNT model can be generalized \cite{Chen:2014ska} by replacing $S^{+}_{2}$ with $\mathbf{\Phi}$ having integer isospin and hypercharge, $Y=1$ and $N_{R_{i}}$ with $\mathbf{F}_{i}$ that has integer isospin and $Y=0$ under SM gauge group. In the generalized KNT model, the lightest neutral fermion component, $F_{1}^{0}$ is the viable DM candidate. Such replacement in KNT model with large electroweak multiplets have been studied for triplet \cite{Ahriche:2014cda}, 5-plet \cite{Ahriche:2014oda} and 7-plet \cite{Ahriche:2015wha} cases. In \cite{Chowdhury:2018nhd}, we have investigated the charged lepton flavor violating processes in the generalized KNT model. In this work we focus on 5-plet and 7-plet cases because the $Z_{2}$ symmetry, $\{S^{+}_{2},N_{R_{i}}\}\rightarrow \{-S^{+}_{2},-N_{R_{i}}\}$ needed to prevent the Dirac neutrino mass term in the Lagrangian, is not required anymore for larger multiplets like in 5-plet and 7-plet cases.

The article is organized as follows. In section \ref{generalizedKNT}, we present the generalized KNT model. Section \ref{SEDM} describes the formalism to calculate Sommerfeld enhanced DM annihilation processes in generalized KNT model. In section \ref{Dmrelic}, we present the relic densities of the DM candidate, including the correction due to SE, in 5-plet and 7-plet cases via thermal freeze-out process and also via non-thermal out-of-equilibrium decay. The DM annihilation cross sections into electroweak bosons at the galactic center are presented in section \ref{result}. Finally we conclude in section \ref{conclusion}.

\section{The Model}\label{generalizedKNT}

Apart from the SM field content, we add the following BSM fields in the generalized KNT model which are charged under SM gauge group, $SU(3)_{c}\times SU(2)_{L}\times U(1)_{Y}$ as
\begin{equation}
\text{Complex scalars:}\,\,\,S^{+}_{1}\sim (0,0,1),\,\,\mathbf{\Phi}\sim (0,j_{\phi},1),\,\,\text{and}\,\,\text{Real fermions:}\,\,\mathbf{F}_{1,2,3}\sim (0,j_{F},0)
\label{particle}
\end{equation}
where $j_{\phi}$ and $j_{F}$ are integer isospin of $SU(2)_{L}$.

In this comparative study, we focus on two set of models in this class;
5-plet model: $\mathbf{\Phi}\sim (0, 2,1)\,\, \&\,\, \mathbf{F}_{1,2,3}\sim (0,2,0)$ and 
7-plet model: $\mathbf{\Phi}\sim (0, 3,1)\,\, \&\,\, \mathbf{F}_{1,2,3}\sim (0,3,0)$.
 
The SM Lagrangian is augmented in the following way,
\begin{equation}
{\cal L}\supset {\cal L}_{SM}+\{f_{\alpha\beta} \overline{L^{c}_{\alpha}}.L_{\beta}S_{1}^{+}
+g_{i \alpha}\overline{\mathbf{F}_{i}}.\mathbf{\Phi}.e_{\alpha_{R}}+h.c\}-\frac{1}{2}\overline{\mathbf{F}^{c}_{i}}M_{F_{ij}}\mathbf{F}_{j}-V(H,\mathbf{\Phi},S_{1})+h.c
\label{eq1}
\end{equation} 
where, c denotes the charge conjugation and dot sign, in shorthand, refers to appropriate $SU(2)$ contractions. Also $L_{\alpha}$ and $e_{R_{\alpha}}$ are the LH lepton doublet and RH charged leptons respectively and Greek alphabet $\alpha$ stands for generation index. Moreover, $[F]_{\alpha\beta}=f_{\alpha\beta}$  and $[G]_{i\alpha}=g_{i\alpha}$ are $3\times 3$ complex antisymmetric and general complex matrices respectively. Finally, $H$ denotes the SM Higgs doublet.

The scalar potential is given by,
\begin{equation}
V(H,\mathbf{\Phi},S_{1})=V(H)+V(\mathbf{\Phi})+V(S_{1})+V_{1}(H,\mathbf{\Phi})+V_{2}(H,S_{1})+V_{3}(\mathbf{\Phi},S_{1})
\label{eq2}
\end{equation}
The three-loop neutrino mass generation and the DM stability depend on the $V_{3}$ term of Eq.(\ref{eq2}). Explicitly the relevant terms of $V_{3}$ for 5-plet and 7-plet models are,
\begin{align}
V^{(5)}_{3}&\supset\frac{\lambda_{S}}{4}(S^{-}_{1})^{2}\mathbf{\Phi}_{abcd}\mathbf{\Phi}_{efgh}\epsilon^{ae}\epsilon^{bf}\epsilon^{cg}
\epsilon^{dh}+\lambda S_{1}^{-}\mathbf{\Phi^{*}}^{abcd}\mathbf{\Phi}_{abef}\mathbf{\Phi}_{cdjl}
\epsilon^{ej}\epsilon^{fl}+h.c\label{eq3}\\
V^{(7)}_{3}&\supset\frac{\lambda_{S}}{4}(S^{-}_{1})^{2}\mathbf{\Phi}_{abcdef}\mathbf{\Phi}_{ghijkl}\epsilon^{ag}\epsilon^{bh}\epsilon^{ci}
\epsilon^{dj}\epsilon^{ek}\epsilon^{fl}+h.c\label{eq4}
\end{align}
Here the $\lambda$ term in Eq.(\ref{eq3}) is not invariant under  $Z_{2}$ and eventually induce the decay of $F^{0}_{1}$ where the width is $\Gamma_{\text{DM}}\sim \lambda^{2}$.  But, as pointed out in \cite{Ahriche:2014oda}, the bound on DM mean life-time sets $\lambda$ to be very small, and  in  the limit when $\lambda\rightarrow 0$, the $Z_{2}$ symmetry emerges. On the other hand, the $\lambda$ term is absent in Eq.(\ref{eq4}) because $j_{\phi}\otimes j_{\phi}$ contains symmetric (antisymmetric) irreducible representation with same isospin $T_{\phi}$ for even (odd) integer isospin which is further contracted with $\mathbf{\Phi}^{\dagger}$ to obtain a singlet. So, for two identical scalar multiplets, the antisymmetric combination is zero and hence no $\lambda$ term for $j_{\phi}=3$.

As pointed out in \cite{Chowdhury:2018nhd}, the mass splittings among component fields of the scalar multiplet is controlled by $\lambda_{H\phi 2}(\mathbf{\Phi}^{\dagger}.H).(H^\dagger.\mathbf{\Phi})\subset V_{2}$ term after electroweak symmetry breaking and allowed splittings only lead to $\Delta m^{2}_{ij}/M_{0}^{2}\sim 10^{-3}$ for invariant mass of the scalar multiplet, $M_{0}=10$ TeV and the ratio becomes smaller for $M_{0}>10$ TeV. On the other hand, the mass splittings in fermionic component fields are zero at tree-level and only receive $O(100)$ MeV splittings due to radiative correction after electroweak symmetry breaking. Therefore such scenario is can be considered as near-degenerate case.

\section{Sommerfeld Enhanced Dark Matter Annihilation}\label{SEDM}

\subsection{DM Candidate}\label{dmcand}

In the generalized KNT model, the lightest neutral component of the fermion multiplet, $F_{1}^{0}$ is the viable DM candidate. In comparison, the neutral component of the scalar multiplet, $\phi^{0}=\frac{1}{\sqrt{2}}(S+iA)$ could have provided $S$ to be DM but it is ruled out as it induces Z-mediated dark matter nucleon scattering of the order $10^{-39}\,\text{cm}^{2}$ which is much larger than the exclusion limit set by the direct detection experiments \cite{Aprile:2017iyp}. One can avoid this  DM-nucleon scattering channel if the splitting between $S$ and $A$ is large enough to make this scattering kinematically forbidden but there is no renormalizable term in the Lagrangian which can induce such splitting in a generic way. Still, higher dimensional operator can split the $S$ and $A$ component \cite{Chowdhury:2016mtl} but then it is needed to address the UV completion of the model. Therefore, we restrict ourselves only to  the renormalizable Lagrangian, and therefore the DM candidate is set to $F_{1}^{0}$.

\subsection{SE Annihilation Cross-sections}\label{secross}

When the DM is non-relativistic, $v_{\text{DM}}\ll c$ and $m_{W,Z}\ll m_{\text{DM}}$, the exchange of massive W and Z  gauge bosons between DM components will induce Yukawa potential and $\gamma$ exchange will induce Coulomb potential which in turn significantly modifies the wavefunction of the incoming DM states and enhances the annihilation cross-sections. This phenomenon is known as Sommerfeld Enhancement (SE). The calculation of Sommerfeld enhanced DM annihilation cross section is well studied subject so here we follow the prescriptions given in \cite{Beneke:2014gja, Chowdhury:2016mtl}. In the following we briefly review them to set up our notation.

As the Sommerfeld enhancement is considered for $2\rightarrow 2$ processes, we first define 2-particle states which consist of incoming component fields of fermion multiplet. Sommerfeld enhancement takes place in DM (co)annihilation processes with final states $W^{\pm}$, $Z$ and $\gamma$ bosons so we only consider the 2-particle states which are CP-even and have total charges $Q=0,\,\pm 1,\,\pm 2$. In the case of DM annihilation in the galaxy halo at present times, only 2-particle states with $Q=0$ are relevant. Moreover, we have considered $M_{F_{1}}<M_{F_{2,3}}$, therefore only component fields of $\mathbf{F_{1}}$ multiplet enter into 2-particle states. We define 2-particle state vector corresponding to $\mathbf{F_{1}}$ as
\begin{align}
Q=0: &\,|\Psi\rangle=(F^{0}_{1}F^{0}_{1},F_{1}^{\pm}F_{1}^{\mp},F_{1}^{\pm\pm}F_{1}^{\mp\mp}, F_{1}^{\pm\pm\pm}F_{1}^{\mp\mp\mp}....)^{T}
\label{se1}\\
Q=\pm 1: &\,|\Psi\rangle=(F_{1}^{0}F^{\pm}_{1},F^{\pm\pm}_{1}F^{\mp}_{1},F_{1}^{\pm\pm\pm}F_{1}^{\mp\mp}...)^{T}\label{se2}\\
Q=\pm 2: &\,|\Psi\rangle=(F_{1}^{0}F^{\pm\pm}_{1},F^{\pm}_{1}F^{\pm}_{1},F_{1}^{\pm\pm\pm}F_{1}^{\mp}...)^{T}\label{se3}
\end{align}
The ordering of the component 2-particle states, as above, is arbitrary. Therefore, one can choose any other ordering of states.

The modification of the wavefunction is determined by solving the radial Schrodinger equation with effective potential,
\begin{equation}
\frac{d^2 \Psi_{jj',ii'}}{d r^2}+\left[\left((m_{F_{1}}v)^2-\frac{l(l+1)}{r^{2}}\right)\delta_{jj',kk'}-m_{F_{1}}V_{jj',kk'}\right]\Psi_{kk',ii'}=0
\label{se4}
\end{equation}
where $r$ is the magnitude of the relative distance between two component fields in their center-of-mass frame, the kinetic energy of the incoming DM states, i.e. $|ii'=F^{0}_{1}F_{1}^{0}\rangle$ is $E=m_{F_{1}}v^2$, The wavefunction $\Psi_{jj',ii'}$ gives the transition amplitude from $|ii'\rangle$ states to $|jj'\rangle$ states in the presence of effective potential, $V$. The double indices $ii'$, $jj'$ and $kk'$ run over the states of the 2-particle state vector defined in Eq.(\ref{se1}).

We primarily focus on the S-wave annihilation so we set $l=0$ and have
\begin{equation}
 \frac{d^{2}\Psi_{jj',ii'}}{d r^2}+\left[k^{2}_{jj'}\delta_{jj',kk'}+
 m_{F_{1}}\left(\frac{f_{jj',kk'}\alpha_{a} 
e^{-n_{a}m_{W}r}}{r}+\frac{Q_{kk'}^{2}\alpha_{\text{em}}}{r}\delta_{jj',kk'}
\right)\right]
 \Psi_{kk',ii'}=0
 \label{schrod2} 
\end{equation}
Here, $k^{2}_{jj'}=m_{S}(m_{S}v^2-d_{jj'})$ is the momentum associated
with the 2-particle state, $|jj'\rangle$
and $d_{jj'}=m_{j}+m_{j'}-2m_{S}$ denotes the mass differences between DM and 
other states of
the multiplet. $Q_{kk'}$ is the electric charge associated with state 
$|kk'\rangle$. Also, $\alpha_{W}=\alpha$ and $n_{W}=1$ for W boson exchange and 
$\alpha_{Z}=
\alpha/\cos^{2}\theta_{W}$ and $n_{Z}=1/\cos\theta_{W}$ for Z boson exchange. Finally, $f_{jj',kk'}$ is the group theoretical factor associated with $SU(2)$.

Now by using dimensionless variables defined as 
$x=\alpha m_{F_{1}} r$, $\epsilon_{\phi}=(m_{W}/m_{F_{1}})/\alpha$,
$\epsilon_{v}=(v/c)/\alpha$ and 
$\epsilon_{d_{ii'}}=\sqrt{d_{ii'}/m_{F_{1}}}/\alpha$, we re-write the
coupled radial Schrodinger equations as
\begin{equation}
\frac{d^{2}\Psi_{jj', ii'}}{dx^2}
+\left[\hat{k}_{jj'}^{2}\delta_{jj',kk'}+\frac{f_{jj',kk'}n_{a}^{2}e^{-n_{a}
\epsilon_{\phi}x}}{x}+\frac{Q_{kk'}^{2}\sin^{2}\theta_{W}}{x}\delta_{jj',kk'}
\right]\Psi_{kk',ii'}=0
\label{schrodeq}
\end{equation}
where the dimensionless momentum, 
$\hat{k}^{2}_{jj'}=\epsilon_{v}^2-\epsilon_{d_{jj'}}^{2}$.

At large $x$, $\Psi_{jj',ii'}$ behaves as $\Psi_{jj',ii'}\sim T_{jj',ii'}e^{i\hat{k}_{jj'}x}$ where $T_{jj',ii'}$ is the transition amplitude provided the effective potential is dominated by Yukawa potential. Now if the annihilation matrix for final state $f$ is given by $\Gamma^{(f)}_{jj',ii'}$, the annihilation cross section is,
\begin{equation}
\sigma_{F_{1}^{0}F_{1}^{0}\rightarrow f}=c(T^{\dagger}.\Gamma^{(f)}.T)_{F^{0}_{1}F^{0}_{1},F^{0}_{1}F^{0}_{1}}
\label{se6}
\end{equation}
where $c=2$ for $|F^{0}_{1}F^{0}_{1}\rangle$ state as it consists of identical fields.

\section{Dark Matter Relic Density}\label{Dmrelic}
The relic density of DM in the universe is measured by Planck Collaboration as $\Omega_{\text{DM}h^{2}}=0.1199\pm 0.0022\,(68\%\,\text{C. L.})$ \cite{Ade:2015xua}. The fermionic DM in the generalized KNT model can achieve this relic density either by the thermal freeze-out process or non-thermal process as we will describe  below.

\subsection{Thermal Freeze-out of DM}\label{dmthermal}
The thermal freeze-out of fermionic DM of the 5-plet and 7-plet, both proceed mainly through 
\begin{itemize}
\item gauge interactions in dominant $S$-wave and sub-dominant $P$ wave channels as the DM is non-relativistic and they are controlled by gauge coupling $g$ and receive non-negligible Sommerfeld enhancement in mainly $S$-wave annihilation cross-sections.
\item yukawa interactions in sub-dominant $P$-wave channels which are controlled by $g_{i\alpha}$ couplings and are less significant because of large gauge annihilation as pointed out in subsequent discussion.
\end{itemize}

For DM in TeV mass range, the thermal freeze-out may take place either in the broken or symmetric phase of the SM depending on its mass. The critical temperature where the cross-over between high-temperature symmetric phase and low-temperature broken phase of the SM takes place is $T_{c}=159\pm 1$ GeV \cite{DOnofrio:2014rug}. But the gauge singlet with mass at the electroweak scale ($\simgt M_{W}$)and coupled with Higgs may change the dynamics of the phase transition. In that case, the phase transition can be first order and critical temperature can be $T_{c}\sim 100$ GeV \cite{AbdusSalam:2013eya}. So considering the freeze-out temperature of DM as $T_{F}\simgt T_{c}\sim 100$ GeV,  the freeze-out condition, $z_{F}=M_{DM}/T_{F}\sim 20-30$ implies that when $M_{DM}\simgt 2-3$ TeV, the DM freezes out in the symmetric phase of the SM.

When the DM freezes out in the broken phase, the calculation of the Sommerfeld enhanced cross-sections of (co)-annihilation processes which enter into the Boltzmann equation, requires solving the Schrodinger matrix equation with effective potential given in Eq.(\ref{se4}). But this calculation is greatly simplified if the freeze-out takes place in the symmetric phase because then $SU(2)_{L}$ is a good symmetry and therefore we can express the two-particle states, which are direct product states, in terms of definite states of $SU(2)$ irreducible representations. Moreover, as $SU(2)$ gauge bosons are massless in this phase, the effective potential is the Coulomb potential and therefore the Sommerfeld enhanced cross-section is
\begin{equation}
\sigma_{SE}=S\sigma_{0}
\label{se6}
\end{equation}
where $\sigma_{0}$ is the perturbative cross-section and $S$ is given by,
\begin{equation}
S(x)=\frac{\pi x}{e^{\pi x}-1},\,\,\,x=\alpha/\beta
\label{se7}
\end{equation}
Here, $\beta=v_{\text{DM}}/c$ and $\alpha$ is the corresponding coupling of the $SU(2)$ group. Moreover, $Y=0$ for $\mathbf{F}_{1}$ so no need to consider $U(1)_{Y}$ contribution. Following \cite{Strumia:2008cf, deSimone:2014pda, ElHedri:2016onc}, the effective potential is,
\begin{equation}
V=\frac{\alpha}{r}T_{R}\otimes T_{\overline{R}}
\label{se8}
\end{equation}
Now $T_{R}\otimes T_{\overline{R}}=\sum\limits_{\oplus Q}\mathbf{Q}$. So if the dimension of representation $R$ is $n$ and that of $\mathbf{Q}$ is $N$, where $N\leq 2n-1$, the potential for total iso-spin is given as,
\begin{equation}
V=(N^{2}+1-2n^{2})\alpha/8r
\label{se9}
\end{equation}

The DM abundance is calculated by solving the following Boltzmann equation \cite{Giudice:2003jh, Cirelli:2005uq},
\begin{equation}
sZHz\frac{dY}{dz}=-2\left(\frac{Y^{2}}{Y^{2}_{\text{eq}}}-1\right)\gamma
\label{se10}
\end{equation}
Here, $z=M_{DM}/T$. $H$ and $s$ are the Hubble rate in the radiation dominated era and the entropy density respectively. They are given by, 
\begin{equation}
H(z)=\sqrt{\frac{\pi^{2}g_{*}(z)}{90}}\frac{M^{2}_{DM}}{M_{\text{pl}}}\frac{1}{z^{2}}\,\,\,\text{and}\,\,\,s(z)=\frac{2 \pi^2 g_{*s}(z)}{45}\frac{M^{3}_{DM}}{z^{3}}
\label{se11}
\end{equation}
where $M_{\text{pl}}=(8\pi G_{N})^{-1/2}=2.44\times 10^{18}$ GeV is the reduced Planck Mass. Also $g_{*}$ and $g_{*s}$ are the total and effective relativistic degrees of freedom respectively. In addition, $Y=n/s$ where $n$ is the number density and $Y_{\text{eq}}$ is value of $Y$ in thermal equilibrium. And hence  $Z=(1-\frac{1}{3}\frac{z}{g_{*s}}\frac{d g_{*s}}{d z})^{-1}$.

The thermal rate of $2\rightarrow 2$ scattering that involves (co)annihilating component fields of $\mathbf{F}_{1}$ into SM fields, $ij\rightarrow a\,b$, at temperature $T$, is denoted by $\gamma$ and given as,
\begin{equation}
\gamma=\frac{T^{4}}{64\pi^{2}}\int_{4M_{DM}^{2}}^{\infty}d s \,s^{1/2}K_{1}\left(\frac{\sqrt{s}}{T}\right)\hat{\sigma}(s)
\label{se12}
\end{equation}
where $K_{1}(x)$ is the modified Bessel function of first kind and 
the reduced cross section, summed over all (co)annihilation channels, is $\hat{\sigma}(s)=2s\lambda(1,M^{2}_{DM}/s,M^{2}_{DM}/s)\sigma(s)$. Also, $\lambda(a,b,c)=(a-b-c)^{2}-4bc$ is the Kallen function.

At the freeze-out, the DM and other component fields are non-relativistic, therefore we can decompose the reduced cross section up to $P$ wave contribution as follows,
\begin{equation}
\hat{\sigma}=c_{s}\beta+c_{p}\beta^{3}
\label{se13}
\end{equation}
where $\beta$ is the DM velocity (here, $c=1$) in the center of mass frame given by $\beta=\sqrt{1-4 M_{DM}^{2}/s}$.

Now as $V$ is in isospin-1 representation, the 2-particle state $VV$ will be in $\mathbf{1}\otimes \overline{\mathbf{1}}=\mathbf{0}\oplus \mathbf{1}\oplus \mathbf{2}$ representations with definite isospin. Therefore we only have to consider $\mathbf{0}$, $\mathbf{1}$ and $\mathbf{2}$ definite representations coming from $\mathbf{2}\otimes \overline{\mathbf{2}}$ and $\mathbf{3}\otimes \overline{\mathbf{3}}$ in the case of 5-plet and 7-plet respectively. Therefore the potentials for definite representations in the case of 5-plet are,
\begin{equation}
V^{(5)}_{\mathbf{0}}=-\frac{6\alpha}{r},\,\,V^{(5)}_{\mathbf{1}}=-\frac{5\alpha}{r},\,\,
V^{(5)}_{\mathbf{2}}=-\frac{3\alpha}{r}
\label{se14}
\end{equation}
whereas, for 7-plet, they are,
\begin{equation}
V^{(7)}_{\mathbf{0}}=-\frac{12\alpha}{r},\,\,V^{(7)}_{\mathbf{1}}=-\frac{11\alpha}{r},\,\,
V^{(7)}_{\mathbf{2}}=-\frac{9\alpha}{r}
\label{se15}
\end{equation}

Finally, we have the Sommerfeld enhanced S-wave coefficients for 5-plet and 7-plet, $c^{(5)}_{s}$ and $c_{s}^{(7)}$ respectively as follows,
\begin{align}
c^{(5)}_{s}&=\frac{g^4}{8\pi}\left(240\times\langle S(-\frac{6\alpha}{\beta})\rangle_{T}+5\times 84\times \langle S(-\frac{3\alpha}{\beta})\rangle_{T}+3\times 125\times \langle S(-\frac{5\alpha}{\beta})\rangle_{T}\right)\label{se16}\\
c^{(7)}_{s}&=\frac{g^4}{8\pi}\left(1344\times\langle S(-\frac{12\alpha}{\beta})\rangle_{T}+5\times 504\times \langle S(-\frac{9\alpha}{\beta})\rangle_{T}+3\times 350 \times\langle S(-\frac{11\alpha}{\beta})\rangle_{T}\right)\label{se16}
\end{align}
where $\langle S(x)\rangle_{T}$ symbolically denotes the result of integration on $S(x)$ in Eq.(\ref{se12}) because it is an implicit function of $s$. Also, the first, second and third terms of above equations denote the $\mathbf{0}$, $\mathbf{2}$ and $\mathbf{1}$ definite representations respectively.  

On the other hand, the gauge contribution to P-wave coefficient, $c_{p}$ is given as
\begin{equation}
c^{(5)}_{p}=\frac{1215 g^{4}}{8\pi}\,\,\,\text{and}\,\,\,c^{(7)}_{p}=\frac{3717 g^4}{4\pi}
\label{se17}
\end{equation}
Also there will be $P$ wave contribution from yukawa terms,
\begin{equation}
c_{p,g}=\sum_{\alpha,\beta} \frac{|g_{1\alpha}|^{2} |g_{1\beta}|^{2} M_{DM}^{4}(M_{DM}^{4}+M_{\phi}^{4})}{6\pi (M_{DM}^{2}+M_{\phi}^{2})^{4}}
\label{se18}
\end{equation}
where the sum is taken over the charged lepton flavors. As the components of the fermion multiplets are exactly degenerate or almost-degenerate  in the symmetric and broken phases respectively, all the (co)annihilation channels contribute equally.

The set of parameters of generalized KNT model which is relevant for DM relic density calculation via thermal freeze-out is, $\{M_{F_{1}},M_{F_{2}},M_{F_{3}},M_{\phi},g_{1\alpha}\}$ apart from the SM gauge couplings of scalar and fermion multiplets. In this analysis we scan over $M_{F_{1}}\in (1,50)$ TeV, $M_{F_{2,3}}\in M_{F_{1}}+(1,10)$ TeV, $M_{\phi}\in M_{F_{1}}+(10,100)$ TeV. Moreover, $g_{1\alpha}$ couplings are chosen so that they satisfy the neutrino constraints as described in \cite{Chowdhury:2018nhd}.
\begin{figure}[h!]
\centerline{\includegraphics[width=10cm]{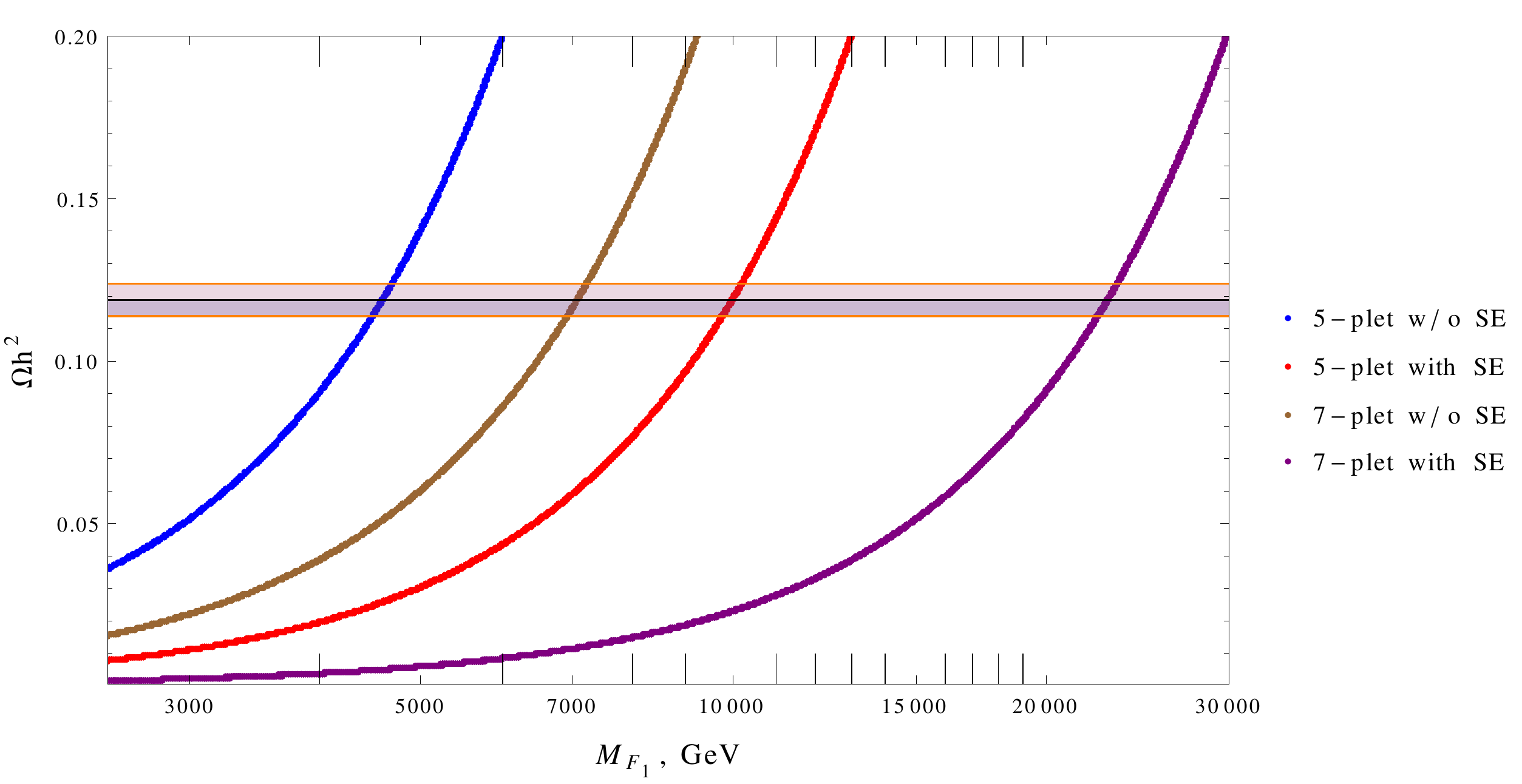}}
\caption{The DM relic densities, $\Omega h^{2}$ of 5-plet w/o SE (blue), 5-plet with SE (red), 7-plet w/o SE (brown) and 7-plet with SE (purple) respectively. The horizontal band represents $5\sigma$ band with central value $\Omega h^{2}=0.1186\pm 0.001$ measured by Planck.}
\label{dmrelic}
\end{figure}

\begin{figure}[h!]
\centerline{\includegraphics[width=8cm]{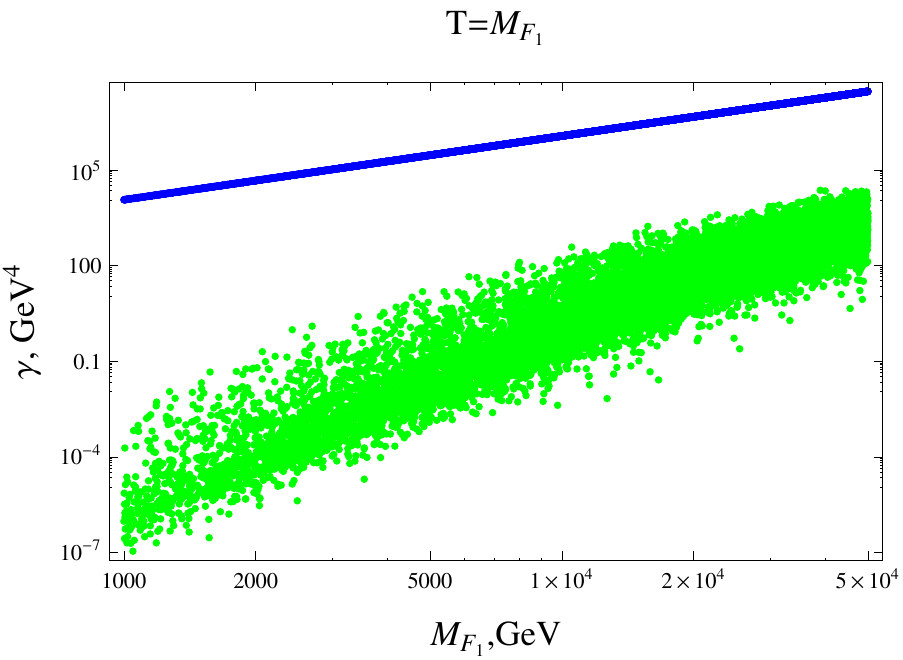}\hspace{0mm}
\includegraphics[width=8cm]{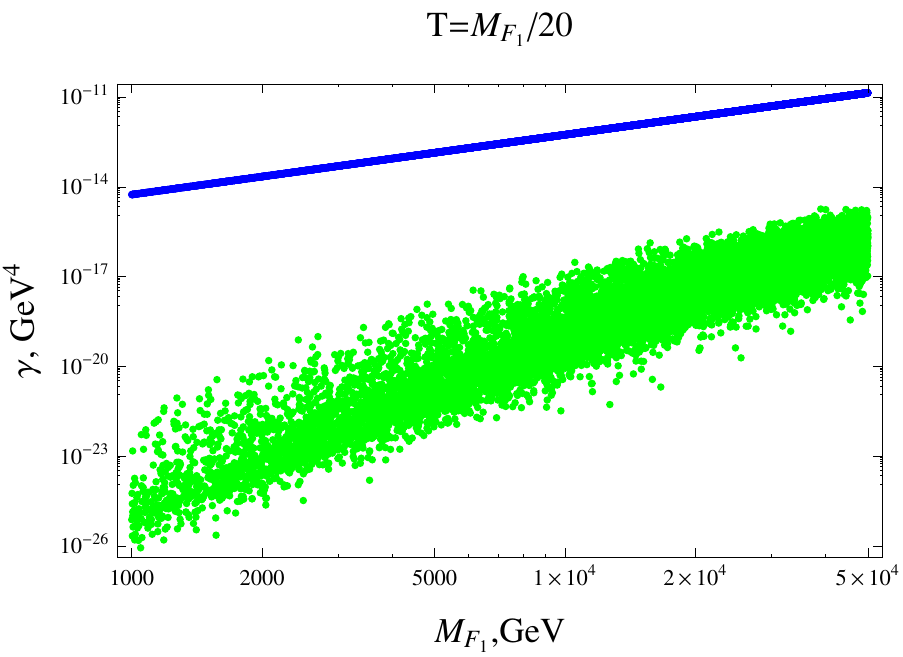}}
\caption{Comparison between the gauge and fermionic contribution in DM freeze-out}
\label{compdm}
\end{figure}

From Fig. \ref{dmrelic}, we can see that the inclusion of Sommerfeld enhanced S-wave contribution significantly changes the mass of the DM for which the correct relic density can be obtained. Moreover, the dominant contribution in thermal freeze-out comes from the gauge contribution as it involves large S-wave and P-wave coefficients in comparison to the P-wave contribution coming from the Yukawa sector of generalized of KNT model, as shown in Fig. \ref{compdm} for two temperature values.

\subsection{Non-thermal Production}\label{nonthermal}
Apart from thermal freeze-out of DM which is mainly controlled by the gauge interactions as seen from Fig. \ref{compdm}, it is possible to set the DM relic  density non-thermally by the out-of-equilibrium decay of $\phi$ scalar via $\phi^{+}\rightarrow F^{0}_{1}e^{+}_{R}$ in generalized KNT model. But as both $\mathbf{\Phi}$ and $\mathbf{F_{1}}$ are charged under the gauge group, the processes $\phi_{i}\phi_{j}\leftrightarrow VV$ and $F_{1_{i}}F_{1_{j}}\leftrightarrow VV$ will keep them in the thermal equilibrium. Therefore, one important condition is that the temperature where the decay takes place must be smaller than the temperatures where the gauge reactions of $\mathbf{\Phi}$ and $\mathbf{F_{1}}$ decouple.

The decay temperature is given as,
\begin{equation}
T_{D}=\left(\frac{90}{\pi g_{*}}\right)^{-1/4}\sqrt{\Gamma_{\phi}M_{pl}}
\label{nonth1}
\end{equation}
where the decay width $\Gamma_{\phi}$  for $\phi^{+}\rightarrow F^{0}_{1}e^{+}_{\alpha}$ process is
\begin{equation}
\Gamma_{\phi}=\sum_{\alpha=e,\mu,\tau}\frac{|g_{1\alpha}|^{2}(M^{2}_{\phi}-M^{2}_{F_{1}})^{2}}{4\pi M^{3}_{\phi}}
\label{nonth11}
\end{equation}

Moreover, the thermal rate of the decay at temperature $T$ is given as
\begin{equation}
\gamma_{D}=\frac{n_{eq}K_{1}\left(M_{\phi}/T\right)\Gamma_{\phi}}{K_{2}\left(M_{\phi}/T\right)}
\label{nonth2}
\end{equation}
where, $n_{eq}$ is the number density of $\phi^{+}$ in equilibrium.

Finally, the gauge reaction density,  $ij\leftrightarrow ab$ at temperature $T$ is given by,
\begin{equation}
\gamma_{A}=\frac{T^{4}}{64\pi^{2}}\int_{s_{\text{min}}}^{\infty}d s \,s^{1/2}K_{1}\left(\frac{\sqrt{s}}{T}\right)\hat{\sigma_{A}}(s)
\label{nonth3}
\end{equation}
where $s_{\text{min}}=\text{Max}\{(M_{i}+M_{j})^{2},(m_{a}+m_{b})^{2}\}$ and $\hat{\sigma}_{A}$ is the reduced $2\rightarrow 2$ scattering cross section of component fields $i,\,j$ of the scalar and fermion multiplets, $\mathbf{\Phi}$ and $\mathbf{F}_{1}$ to SM fields $a,\,b$ via gauge interactions. The decoupling conditions for both scattering and decay rates are set to be $\frac{\gamma}{n_{\text{eq}}H}\leq 1$.

\begin{figure}[h!]
\centerline{\includegraphics[width=10cm]{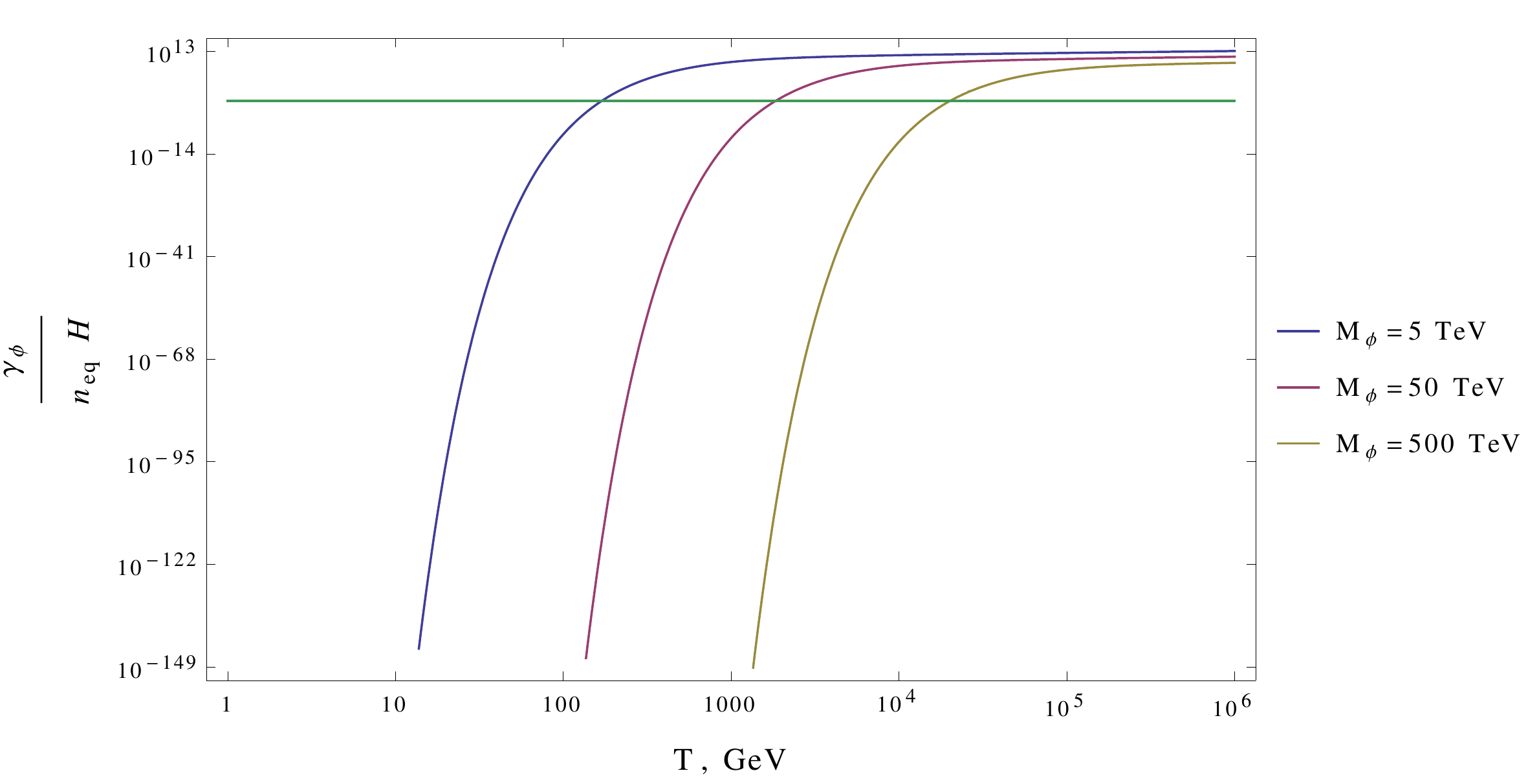}}
\caption{Decoupling of the gauge reaction densities of $\mathbf{\Phi}$ with temperature.}
\label{gaugerec}
\end{figure}
\begin{figure}[h!]
\centerline{\includegraphics[width=12cm]{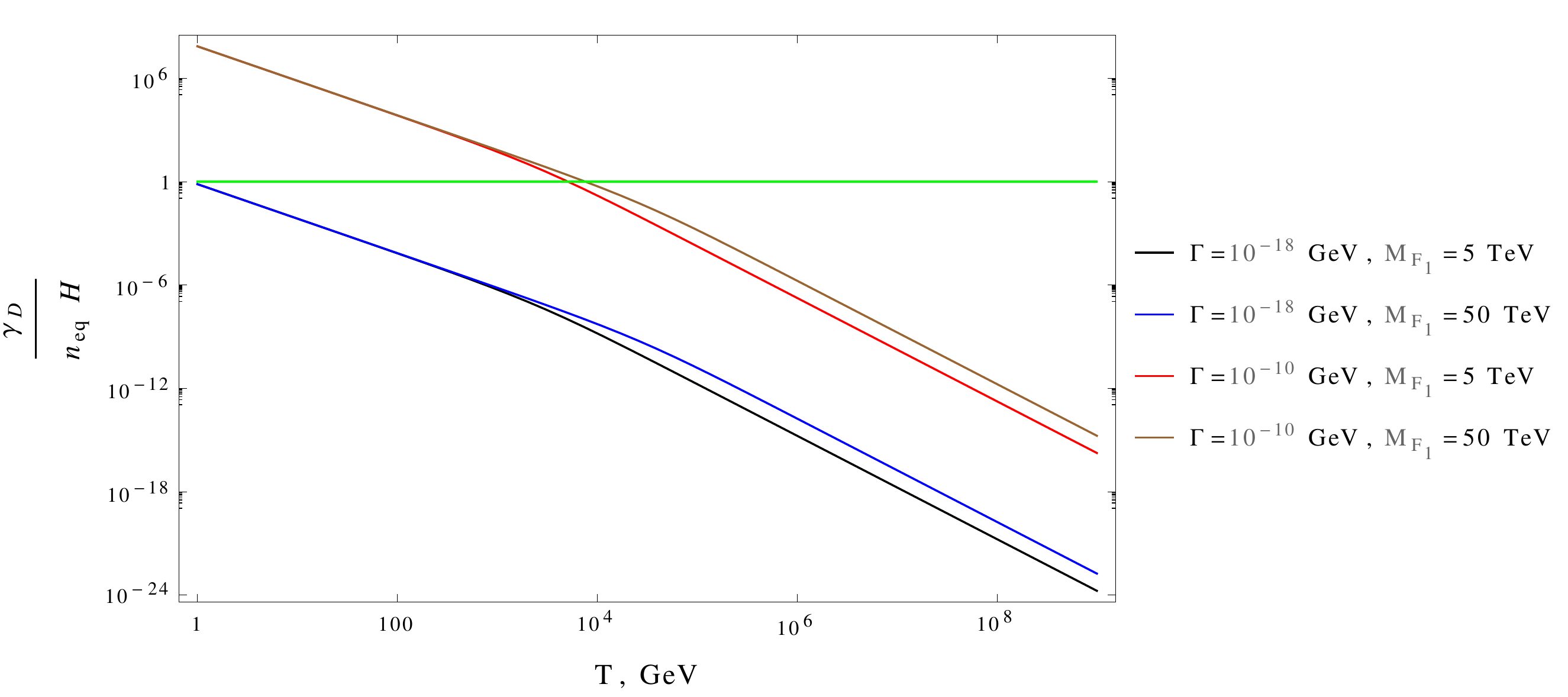}\hspace{0mm}\includegraphics[width=5.5cm]{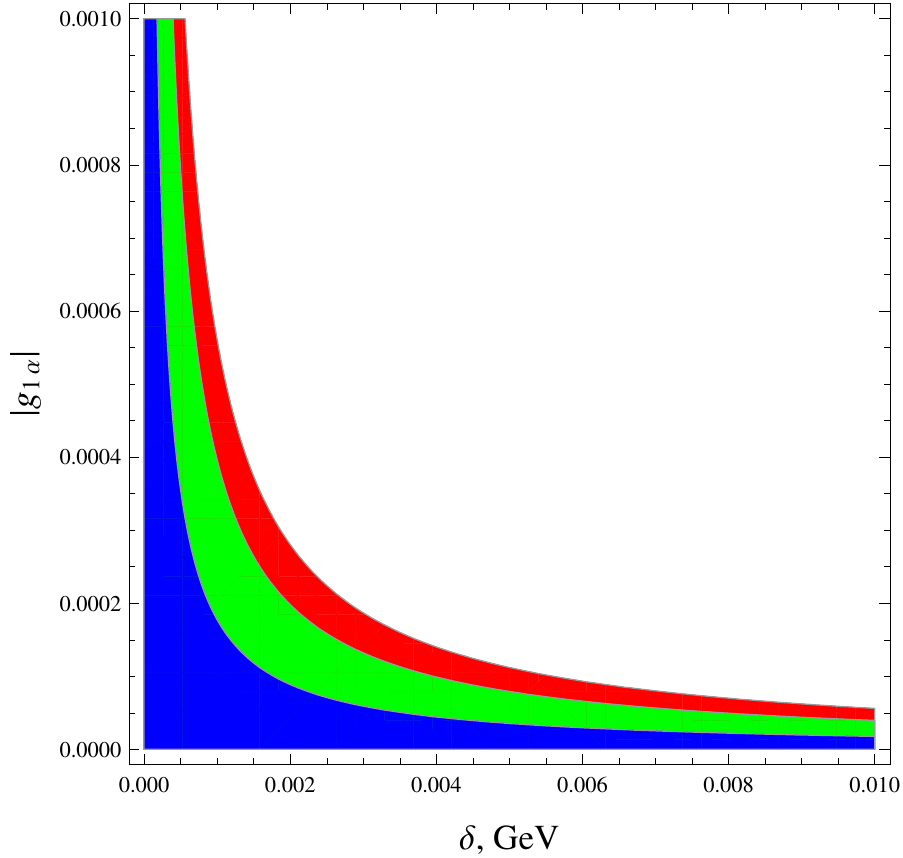}}
\caption{(left) Decoupling of inverse decay process with temperature and (right) Allowed region on $\delta-|g_{1\alpha}|$ plane due to $\Gamma_{\phi}\leq 10^{-18}$ GeV for $M_{\phi}=10$ TeV (blue), $M_{\phi}=50$ TeV (green), $M_{\phi}=100$ TeV (red) respectively.}
\label{invdec}
\end{figure}

From Fig. \ref{gaugerec}, we can see that the gauge reaction densities of $\mathbf{\Phi}$ component fields decouple when temperature becomes small. One the other hand, the decoupling of the inverse decay process $F_{1}^{0}e^{+}_{R}\rightarrow \phi^{+}$ which would deplete the amount of $F_{1}^{0}$, is necessary. This condition sets the corresponding decay width to be very small, at the order of $\sim 10^{-18}$ GeV so that the inverse process remains decoupled throughout the whole thermal history of the universe as shown in the Fig. \ref{invdec} (left).

Besides, Fig. \ref{invdec} (right) represents the $\delta-|g_{1\alpha}|$ plane bounded by the constraint $\Gamma_{\phi}\simlt 10^{-18}$ GeV so that the inverse decay process remains out of equilibrium during the thermal evolution of the universe. We can see that, such small decay width of $\phi^{+}$ implies that the mass difference between $\phi^{+}$ and $F_{0}^{1}$ needs to be of  the order $\mathcal{O}(1-10)$ MeV and $|g_{1\alpha}|\sim 10^{-4}$ for $M_{\phi}\sim 10$ TeV. Therefore from this estimates, we can infer that the out-of-equilibrium decay of $\phi^{+}$ to generate DM content of the universe only holds for a fine-tuned parameters of the model. Nevertheless, one can extend the generalized KNT model with another sector which can non-thermally produce the DM without any fine-tuning. The detailed construction of such extended model and its connection to the thermal history of the universe are beyond the scope of this paper and left for future investigation.

\section{Result and Discussion}\label{result}

In section \ref{dmrelic}, we have identified the masses of $F_{0}^{1}$ for which standard thermal freeze-out process gives the current relic density in both 5-plet and 7-plet cases. Moreover, if we relax the thermal freeze-out scenario and consider the non-thermal production of DM in the generalized KNT model, one can achieve the current DM density for a wide range of masses, $5-50$ TeV but to achieve that,  fine-tunings in mass difference $|M_{\phi}-M_{F_{1}}|$ and $|g_{1\alpha}$ are required. Still one can have extended KNT model with a dark sector which can assist the non-thermal DM production but will not be subject to any fine-tuning. Therefore, in the subsequent analysis of Sommerfeld enhanced DM annihilation, we have considered the range $1-50$ TeV because this range has better sensitivity in IACTs.

\subsection{DM Direct Detection}\label{directdet}
The DM candidate $F^{0}_{1}$ does not couple to quarks at tree-level because of its vanishing hypercharge. However at one loop level, due to exchange of $W$ boson, it has effective coupling with the quarks which leads to both spin-dependent and spin-independent contribution in DM-nucleon scattering. The spin-dependent cross-section is suppressed by the mass of the DM which is at O(TeV). On the other hand, the spin-independent cross-section for fermionic multiplet with integer isospin $j$, which doesn't depend on the DM mass, is given by
\cite{Cirelli:2005uq},
\begin{equation}
\sigma_{\text{SI}}=
j^2(j+1)^{2}\frac{\pi \alpha^2 M_{\text{Nucl}}^{4}f^{2}}{4 m_{W}^{2}}\left(\frac{1}{m_{W}^{2}}+\frac{1}{m_{h}^{2}}\right)
\label{spinind}
\end{equation}
where, $M_{\text{Nucl}}$ is the mass of the target nucleus, $f$ parametrizes nucleon matrix element as $\langle n|\sum_{q}m_{q}\overline{q}q|n\rangle=f\,m_{n}\overline{n}n$ and from lattice result, $f=0.347131$ \cite{Giedt:2009mr}.

\begin{figure}[h!]
\centerline{\includegraphics[width=12cm]{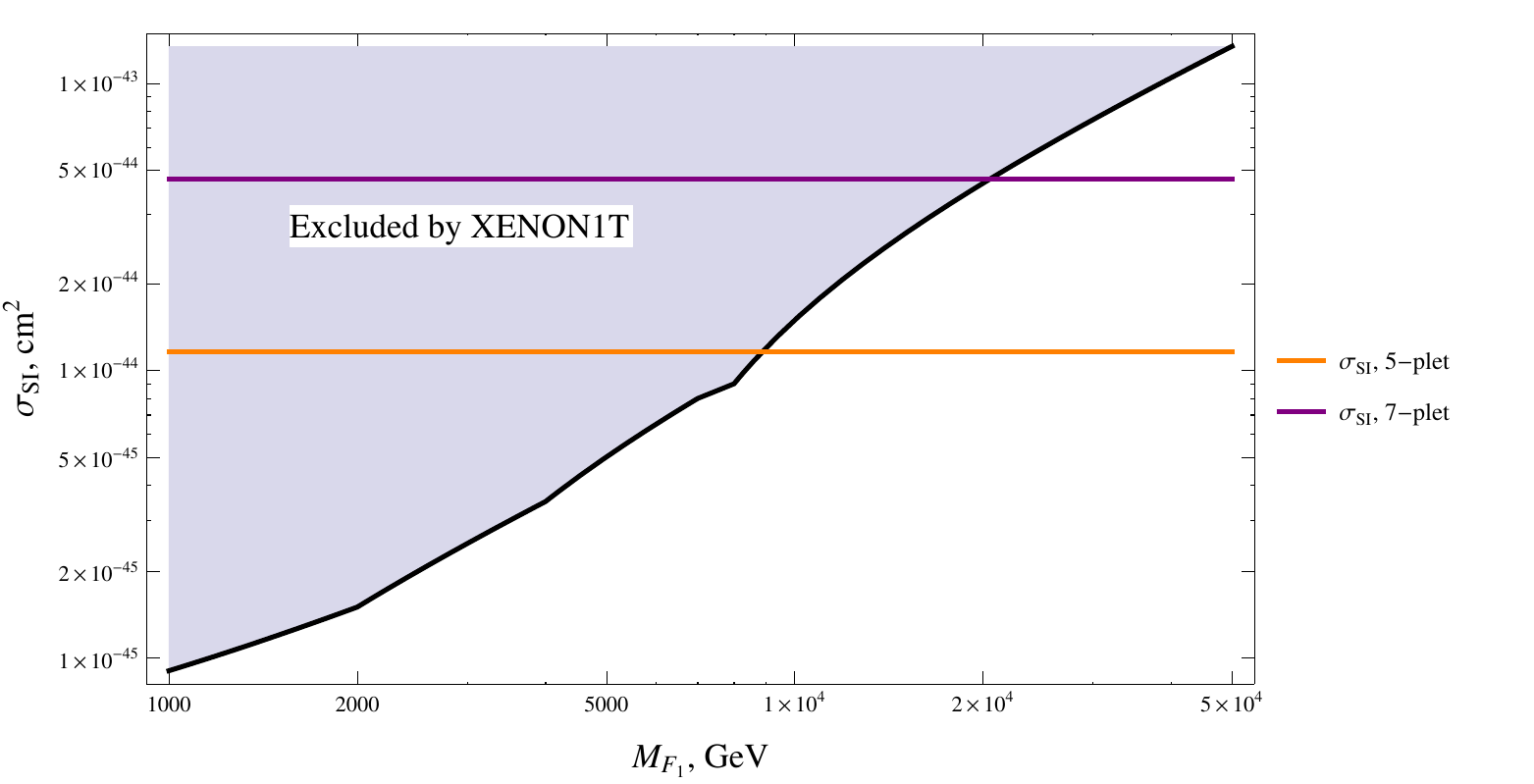}}
\caption{Spin-independent cross section of DM-nucleon interaction. The shaded region is excluded by XENON1T (2017) data \cite{Aprile:2017iyp}. The exclusion limit on DM mass in \cite{Aprile:2017iyp} is given up to 10 TeV. Here we have extrapolated this exclusion limit up to 50 TeV. As we can see, the thermal DM scenario with $M_{F_{1}}=9.9$ TeV for 5-plet and $M_{F_{1}}=22.85$ TeV for 7-plet, are almost at the verge of exclusion by the XENON1T (2017).}
\label{dmdirect}
\end{figure}

\subsection{Sommerfeld Enhanced Cross-sections}\label{secrossdis}
In this section we present the Sommerfeld enhanced cross-sections of  $F^{0}_{1}F^{0}_{1}\rightarrow W^{+}W^{-}$ and $F^{0}_{1}F^{0}_{1}\rightarrow \gamma\gamma$ which are sensitive to IACTs. Moreover, in the analysis we set, $v_{\text{DM}}=10^{-3}$ which is the scale of DM average velocity in the galactic halo.

\begin{figure}[h!]
\centerline{\includegraphics[width=12cm]{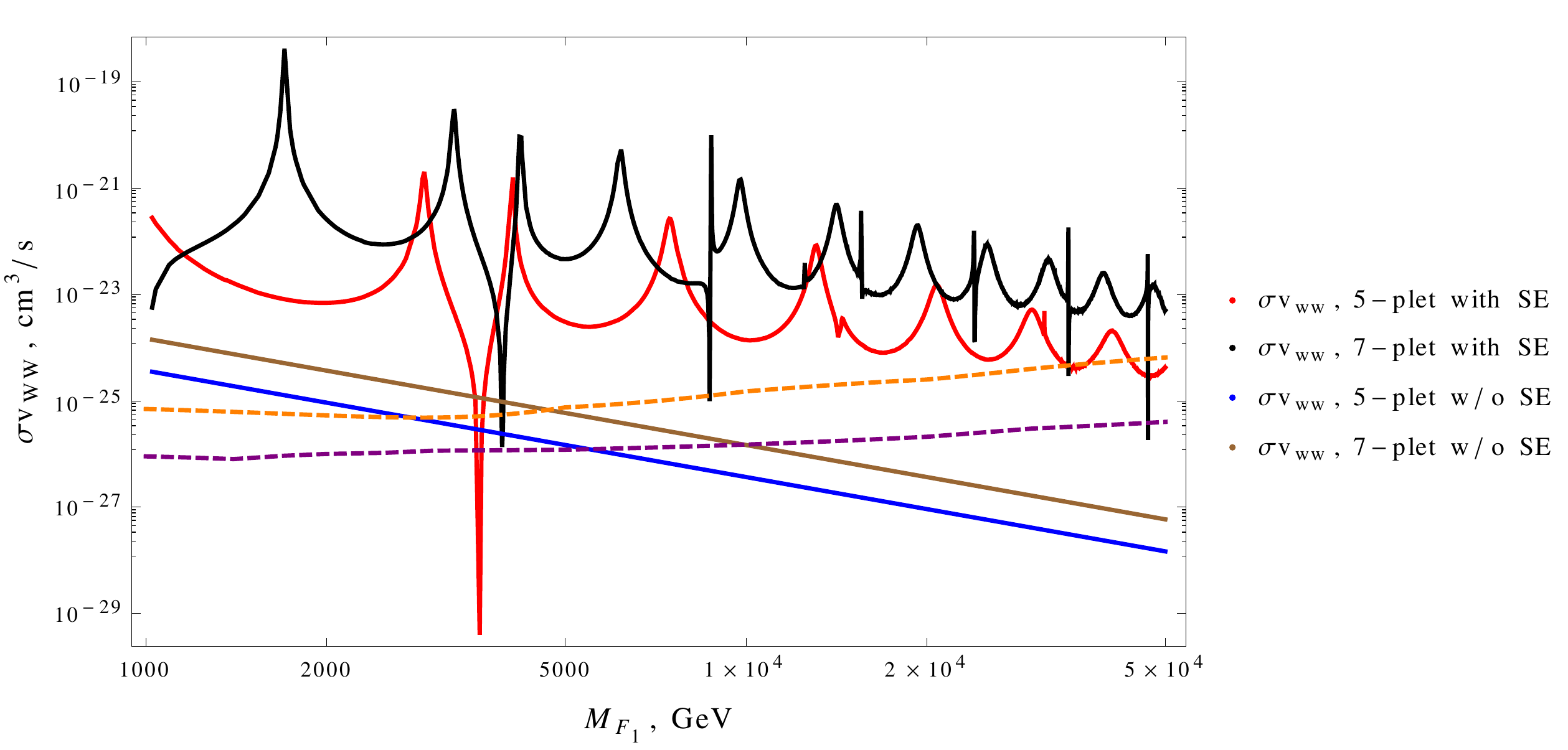}}
\caption{Sommerfeld enhanced cross-section $\sigma v_{\text{WW}}$ at the galactic halo of the Milky Way for 5-plet (red) and 7-plet (black). Moreover, $\sigma v_{\text{ww}}$ without SE is given for 5-plet (blue) and 7-plet (brown). Here orange and purple dashed lines are H.E.S.S observed limit and CTA sensitivity limit on $\sigma v_{\text{ww}}$ respectively}
\label{secross1}
\end{figure}

In Fig. \ref{secross1}, we can see the resonance and dips occurring for both 5-plet (red) and 7-plet (black) at particular mass values of DM due to SE in the presence of Yukawa potential induced by the exchange of massive $W$ and $Z$ bosons in the limit of non-relativistic velocity. Apart from the dips at 3.8 TeV (5-plet) and 4 TeV (7-plet), $\sigma v_{\text{ww}}$ is larger than its tree-level value (blue and brown lines for 5-plet and 7-plet respectively) for almost all of the DM mass range, 1-50 TeV. In fact for this mass range, it is large enough to be almost excluded by the H.E.S.S. limit (orange dashed line) provided that the $F_{1}^{0}$ is the dominant DM of the universe and follows the Einasto density profile \cite{::2016jja}. In addition we can see from Fig. \ref{secross1} that the future CTA will improve the exclusion limit by a factor of $\mathcal{O}{(10)}$ (purple dashed line) \cite{Acharya:2017ttl}.  

\begin{figure}[h!]
\centerline{\includegraphics[width=12cm]{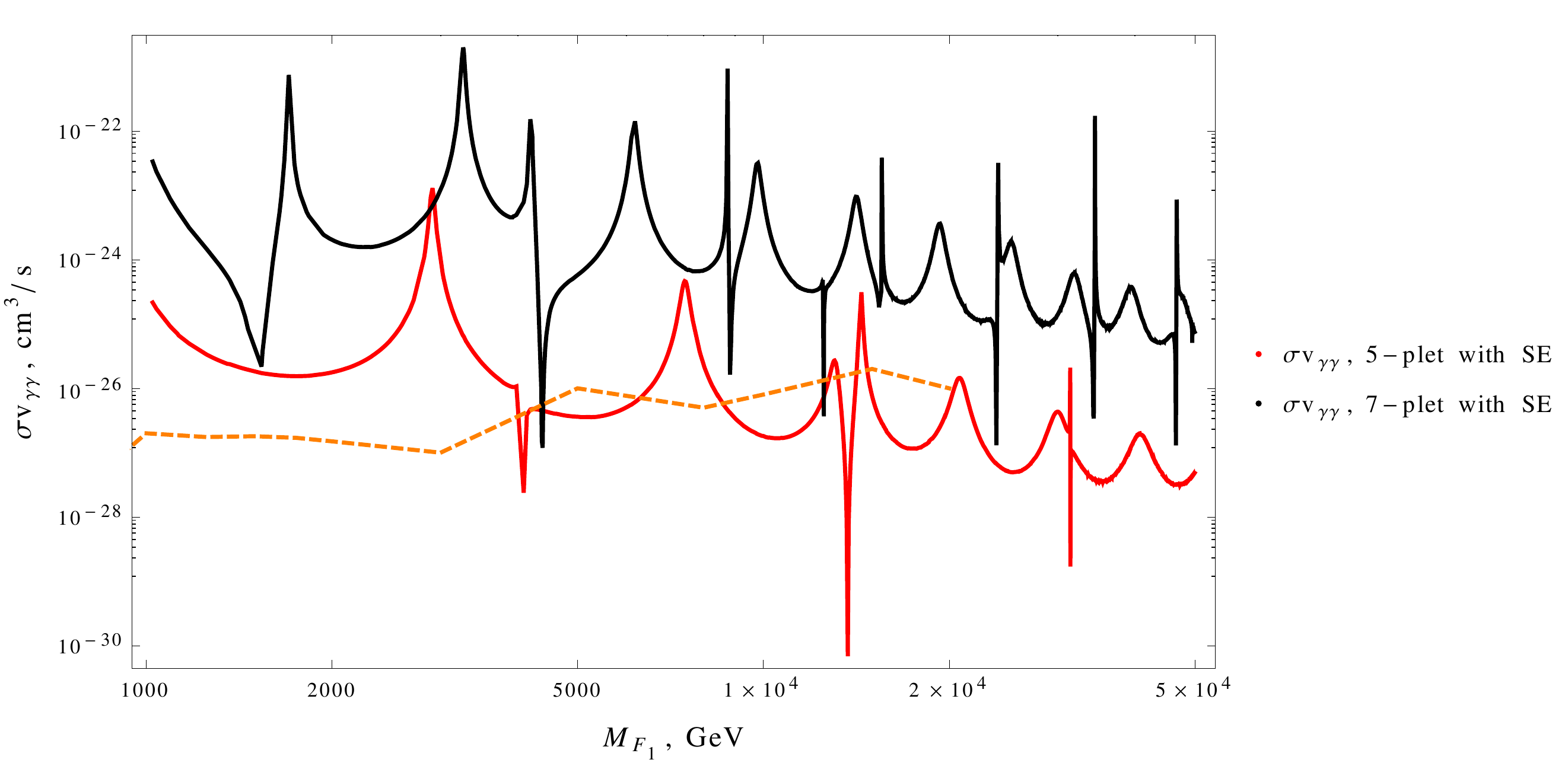}}
\caption{Sommerfeld enhanced $\sigma v_{\gamma\gamma}$  at the galactic halo of the Milky Way for 5-plet (red) and 7-plet (black). Here the orange dashed line is H.E.S.S. observed limit on $\sigma v_{\gamma\gamma}$.}
\label{secross2}
\end{figure}

In addition, Fig. \ref{secross2} represents the Sommerfeld enhanced cross section, $\sigma v_{\gamma\gamma}$ for the process, $F_{1}^{0}F^{0}_{1}\rightarrow\gamma\gamma$. At tree-level, this process does not take place because the DM is charge neutral but due to the multiple exchange of gauge bosons i.e $W^{\pm}$, $Z$ and $\gamma$ and charged states, $F^{(\pm Q)}_{1}$ in the ladder diagrams, the effective coupling with the photons is possible when the DM is non-relativistic. If $\epsilon_{\phi}\simgt 1$ and/or $\epsilon_{v}\simgt 1$, the Sommerfeld enhancement will be suppressed and in that case, $F_{1}^{0}F^{0}_{1}\rightarrow\gamma\gamma$ proceeds through one-loop process that gives, for 1-50 TeV mass range, $\sigma v_{\gamma\gamma}$ of the order $10^{-28}-10^{-31}\,\text{cm}^{3}\text{s}^{-1}$. Again, we can see from Fig. \ref{secross2} that apart from some dips, for almost all of 1-20 TeV mass range, H.E.S.S. (orange dashed line) can exclude the DM in case of 5-plet (red line) and 7-plet (black line) using gamma-line searches.

In passing, we would like to point out that, unlike the case of scalar DM with larger electroweak multiplet which is focused in \cite{Chowdhury:2016mtl}, the mass splittings among the fermionic component fields do not suppress the Sommerfeld enhancement as they are nearly degenerate for O(TeV) mass range.

\section{Conclusion and Outlook}\label{conclusion}

In this work, we have presented the Sommerfeld enhanced DM annihilation cross sections to gamma rays for a class of  three loop radiative neutrino mass generation models  with large electroweak fermion and scalar multiplets i.e 5-plet and 7-plet under $SU(2)_{L}$. We have shown that larger multiplet leads to more enhanced annihilation cross section because more charged component fields of the fermion multiplet contribute to the ladder diagrams. As a consequence, the potential matrix element, $V_{ii',jj'}$ induced by $W$ exchange depends roughly on the group theoretical factor, $j(j+1)-m(m+1)$ where $j$ is the total isospin and $m$ is the $T_{3}$ value of the component field in the multiplet. Moreover, the annihilation matrix element into final states $WW$, $\Gamma^{(WW)}_{ii',jj'}$ also depends on factor, $j(j+1)-m^{2}$. Therefore, larger multiplets will naturally give rise to larger cross sections that we have already observed by comparing $\sigma v_{\text{ww}}$ and $\sigma v_{\gamma\gamma}$ for 5-plet and 7-plet in Fig. \ref{secross1} and Fig. \ref{secross2} respectively.

Finally, putting astrophysical uncertainties aside, because of the larger Sommerfeld enhancement for the higher electroweak multiplets, we can see from Fig. \ref{secross1} and Fig. \ref{secross2} that for almost all of mass range, 1-50 TeV, the constraints from H.E.S.S. can exclude the $F^{0}_{1}$ being the DM, provided it is the dominant DM component and follows the Einasto density profile in both cases of 5-plet and 7-plet. That leaves only the singlet and triplet case as a viable DM candidate in the generalized KNT model. The singlet fermion, $N_{R_{1}}$ of the KNT model which is electroweak neutral, does not receive any Sommerfeld enhancement. On the other hand, the triplet case, where the DM candidate is the neutral component of the fermion multiplet with isospin, $j=1$, will have enhanced annihilation processes due to exchange of electroweak bosons but being smaller representation than the 5-plet or 7-plet, it may have larger potion of parameter space yet to be excluded by the H.E.S.S. limit but will be within the reach of future CTA sensitivity limits. The detailed analysis for gamma rays coming from DM annihilation in different astrophysical environments (galactic center and dwarf spheroidal galaxies) for the triplet case is left for future investigation \cite{chowdhury1}. 

\subsection*{Acknowledgement}
T.A.C. would like to thank Shaaban Khalil for the invitation to the {\it Beyond Standard Model: From Theory to Experiment} held on 17-21 December 2017 in Hurghada, Egypt where the preliminary results of this work were presented.

\end{document}